\documentstyle[preprint,aps]{revtex}

\begin{document}
\draft
\tighten

\preprint{\vbox{\hbox{CLNS 94/1280 \hfill}
                \hbox{CLEO 94--10  \hfill}
                \hbox{\today       \hfill}}}

{
\title{\Large \bf Production and Decay of D$_1(2420)^0$ and D$_2^*(2460)^0$ }

\author{
P.~Avery,$^{1}$ A.~Freyberger,$^{1}$ J.~Rodriguez,$^{1}$
R.~Stephens,$^{1}$ S.~Yang,$^{1}$ J.~Yelton,$^{1}$
D.~Cinabro,$^{2}$ S.~Henderson,$^{2}$ T.~Liu,$^{2}$ M.~Saulnier,$^{2}$
% ...
}

\address{
\bigskip 
{\rm (CLEO Collaboration)}\\  
\newpage 
$^{1}${University of Florida, Gainesville, Florida 32611}\\
$^{2}${Harvard University, Cambridge, Massachusetts 02138}\\
% ...
\bigskip 
}        

\maketitle
}

\begin{abstract}
We have investigated $D^{+}\pi^{-}$ and $D^{*+}\pi^{-}$ final states
and observed the two established $L=1$ charmed mesons, the $D_1(2420)^0$ with
mass $2421^{+1+2}_{-2-2}$ MeV/c$^{2}$ and width $20^{+6+3}_{-5-3}$ MeV/c$^{2}$
and the $D_2^*(2460)^0$ with mass $2465 \pm 3 \pm 3$ MeV/c$^{2}$ and width
$28^{+8+6}_{-7-6}$ MeV/c$^{2}$. Properties of these final states,
including their decay angular distributions and spin-parity
assignments, have been studied.
We identify these two mesons as the $j_{light}=3/2$ doublet predicted by HQET.
We also obtain constraints on {\footnotesize $\Gamma_S/(\Gamma_S + \Gamma_D)$}
as a function of the cosine of the relative phase of the two amplitudes in the
$D_1(2420)^0$ decay.
\end{abstract}

\newpage

\section{Introduction}

Heavy Quark Effective Theory (HQET) predicts the presence of an
approximate flavor-spin symmetry for hadrons containing one heavy quark
($m_{Q} \gg \Lambda_{QCD}$) \cite{r01,r02}.
One of the outstanding issues in this theory is whether the charm quark
is sufficiently heavy for the approximations made in the theory to be
valid. One testable prediction of HQET is the partial wave structure
of the decays of the $D_J$ mesons.
The $D_J$ mesons consist of one charmed quark ($Q$) and one light quark
($\overline{q}$) with relative orbital angular momentum $L$. When $L=1$,
there are four states with spin-parity $J^{P} = 0^{+}$, $1^{+}$, $1^{+}$
and $2^{+}$. In the notation introduced by the Particle Data Group \cite{r03},
these states are labeled $D_{0}^{*}$, $D_{1}$ for both $1^{+}$ states,
and $D_{2}^{*}$, respectively.

Parity and angular momentum conservation place restrictions on the strong
decays of these $D_J$ states to $D \pi$ and $D^{*} \pi$:
the $0^+$ state can decay only to $D \pi$ through S-wave decay,
either $1^+$ state can decay to $D^* \pi$ through S-wave or D-wave decays,
and the $2^+$ state can decay to both $D \pi$ and $D^* \pi$ only through
D-wave decays.

In the decay $D_J \rightarrow D^{*}\pi$, the helicity angular
distribution of the $D^{*}$ can be used to analyze the spin of the parent
$D_J$. The helicity angle, denoted by $\alpha$, is defined as the angle
between the
$\pi^{-}$ from the decay $D_J^0 \rightarrow D^{*+}\pi^{-}$ and the
$\pi^{+}$ from the decay $D^{*+} \rightarrow D^{0}\pi^{+}$, both measured
in the $D^{*+}$ rest frame. Regardless of the initial
polarization of the $D_J$ states, the predicted helicity angular
distributions are:
\begin{equation}
 \frac{d \ N}{d \ \cos\alpha} \propto \left\{ \begin{array}{ll}
  \sin^{2}\alpha & \mbox{($2^{+}$ state)}  \\
  1              & \mbox{(pure S-wave $1^{+}$ state)}\\
  1 \; + \; 3\cos^{2}\alpha & \mbox{(pure D-wave $1^{+}$ state})
                  \end{array} \right.
\end{equation}

Two $D_J^0$ states have been observed previously \cite{r04,r05,r06,r07,r08}.
However, the decay angular analyses performed were all incomplete and
not in total agreement with each other.

In the limit $m_{Q} \rightarrow \infty$, the mesons
are described by $j = S_{q} + L$ (total angular momentum of the light quark).
When $m_{Q}$ is large, but not infinite, the properties of the mesons will also
depend on $S_{Q}$ (spin of the heavy quark) and $J = j + S_{Q}$
(total angular momentum). The mesons with $L = 1$ are then labeled as:
\begin{equation}
 L_{J}^{(j)P} = \left\{ \begin{array}{ll}
     P_{0}^{(1/2)+} \; , \; P_{1}^{(1/2)+}  & \; \; \; (j=1/2 \ \rm{doublet})\\
     P_{1}^{(3/2)+} \; , \; P_{2}^{(3/2)+}  & \; \; \; (j=3/2 \ \rm{doublet})
                  \end{array} \right.
\end{equation}

The $j=1/2$ mesons are predicted to decay exclusively in an S-wave,
while the $j=3/2$ mesons decay only in a D-wave \cite{r01,r09}.
Mesons which decay via a D-wave
are predicted to  be relatively narrow (widths of tens of MeV/c$^{2}$),
while those of the other doublet are predicted to be quite broad
(hundreds of MeV/c$^{2}$). The $P_{2}^{(3/2)+}$ state can decay to both
$D\pi$ and $D^{*}\pi$. Models \cite{r09,r10,r11} predict the ratio of the
branching fractions, $B(D_2^*\rightarrow D\pi)/B(D_2^*\rightarrow D^{*}\pi)$,
to lie in the range from 1.5 to 3.0.

Because the mass of the charmed quark is not infinite, the
$P_{1}^{(3/2)+} \rightarrow D^*\pi$ decay
can also proceed via an S-wave. HQET predicts that this S-wave amplitude is
small compared with typical S-wave amplitudes, but makes no prediction
for the relative magnitudes of the S-wave and D-wave amplitudes
for this decay. However, particular models do make such predictions
\cite{r10,r11}.

The large data sample collected with the CLEO II detector allows us to
identify clearly two of the $D_J^0$ states, to measure their
corresponding angular distributions and to test some of the HQET
predictions.

\section{Data Sample and Event Selection}

The data used in this analysis
were selected from hadronic events produced
in $e^{+}e^{-}$ annihilations at CESR and collected with the CLEO II detector.
Both neutral and charged particles are measured with excellent
resolution and efficiency by the CLEO II detector. A detailed
description of the detector can be found elsewhere \cite{r12}.
The center-of-mass energies used in this analysis were at the mass of the
$\Upsilon(4S)$, $E_{C.M.} = 10.580$ GeV, and in the nearby continuum.
The data corresponds to an integrated luminosity of 1.7 fb$^{-1}$.

We selected events that  have a minimum of three charged
tracks, a total visible energy greater than 15\% of the center-of-mass
energy (this reduced contamination from two-photon interactions and beam-gas
events), and a primary vertex within $\pm 5$ cm in the z-direction
and $\pm 2$ cm in the r-$\phi$ plane of the nominal collision point.
All charged tracks were required to have dE/dx information. If available,
time-of-flight information was also used.

When reconstructing $\pi^{0}$ candidates, we used pairs of photons from the
barrel region with $|\cos\theta| < 0.707$, where the energy resolution is
best. The photons were required to have a minimum energy of 50 MeV
and to be isolated from charged tracks.
All $\pi^{0}$ candidates, whose invariant mass was within 50 MeV/c$^2$ of
the $\pi^{0}$ mass, were kinematically fit to the known $\pi^{0}$ mass and
were required to have a minimum momentum of 65 MeV/c.

\section{$\bf D_2^{*0} \rightarrow D^{+}\pi^{-}$}

To reconstruct $D_2^{*0} \rightarrow D^{+}\pi^{-}$ \footnote[3]{References in
this paper to a specific state or decay will always imply  that the
charge-conjugate state or decay has been included as well.},
we first reconstructed the $D^{+}$ in the decay mode
$K^{-}\pi^{+}\pi^{+}$. We required that the decay angle $\theta_{K}$, the
angle between the direction of the $D^{+}$ momentum in the lab and the
direction of the
$K^{-}$ momentum in the $D^{+}$ rest frame, satisfy the
condition $\cos\theta_{K} < 0.8$, since the background peaks near 1.
Each $D^{+}$ candidate was then combined with each remaining $\pi^{-}$ in
the event. In order to reduce combinatorial background we applied the cuts of
$x_{p}(D_2^{*0})=p(D_2^{*0})/
\mbox{\footnotesize $\sqrt{E_{beam}^{2}-M(D_2^{*0})^{2}}$}>0.65$
and $\cos\theta_{\pi}>-0.8$, where the decay angle, $\theta_{\pi}$,
is defined as the angle between the direction of the $D_2^{*0}$ momentum
in the lab and the direction of the $\pi^{-}$ momentum in the $D_2^{*0}$ rest
frame.
We then calculated the total probability,
$P_{tot}$, of the candidate using the particle identification
(dE/dx and time-of-flight) and the
reconstructed $D^{+}$ mass. $P_{tot}(\chi^{2}_{tot},N_{dof})$
is defined as the probability to observe $\chi^{2}>\chi^{2}_{tot}$ for
$N_{dof}$ degrees of freedom.
The data sample is highly contaminated due to the small $D^+$ signal to
background ratio, and this produces a large and broad peaking in
the $P_{tot}$ distribution at $P_{tot} = 0$.
For the signal this distribution is expected to be flat. We accordingly
required $P_{tot} > 0.4$.

The spectrum of the mass-difference, $M(D^{+}\pi^{-})-M(D^{+})$, for all
$D^{+}\pi^{-}$ combinations surviving the above cuts is shown
in Fig.\ \ref{figm2}. This spectrum was fitted with a third-order
Chebychev polynomial for the background and a Breit-Wigner resonance shape,
convoluted with a
Gaussian resolution function, for the signal. The $\sigma$ of this Gaussian
function was fixed to 4.5 MeV/c$^{2}$, as determined from Monte Carlo studies.
The region from 380 to 430 MeV/c$^2$  was excluded from the fit because
this region is populated by feed-down, caused by not reconstructing
neutrals in the decay chain, $D_J^0 \rightarrow D^{*+}\pi^{-}$, with
$D^{*+} \rightarrow D^{+}\pi^{0}$ or $D^{+}\gamma$.
Our fit yielded $486^{+103}_{-119} $ signal events with a value
$M(D_2^{*0})-M(D^{+})= 596 \pm 3 \pm 3$ MeV/c$^{2}$, which corresponds to a
$D_2^{*0}$ mass of  $2465 \pm 3 \pm 3$ MeV/c$^{2}$, and an intrinsic width
$\Gamma=28^{+8+6}_{-7-6}$ MeV/c$^{2}$. The second error is systematic and was
estimated by varying the cuts, the background parameterization, and the spin
of the Breit-Wigner distribution used. Our results for the mass and
width of this state, along with previous measurements, are listed in
Table \ref{tabm2}.

\section{$\bf D_J^{0} \rightarrow D^{*+}\pi^{-}$}

To reconstruct $D_J^0 \rightarrow D^{*+}\pi^{-}$ we first reconstructed
$D^{0}$'s in the decay
modes $K^{-}\pi^{+}$, $K^{-}\pi^{+}\pi^{0}$, and $K^{-}\pi^{+}\pi^{+}\pi^{-}$.
The $D^{*+}$ candidates were reconstructed by
combining each $D^{0}$ candidate with each remaining $\pi^{+}$ in the event.
Each $D^{*+}$ candidate was then combined with each remaining
$\pi^{-}$ in the event.
In order to reduce combinatorial background, a cut of $x_{p}(D_J^0)>0.6$
and a cut of $\cos\theta_{\pi}>-0.7$ were applied.
We calculated  $P_{tot}$ using the particle identification
(dE/dx and time-of-flight), the $D^{0}$ mass,
the $\pi^0$ mass, and the mass-difference $M(D^{*+})-M(D^{0})$. A
purifying cut of $P_{tot}>0.05$ was then imposed.

The expected angular distributions, outlined previously, can be used to
separate the two $D_J^0$ states decaying
to $D^{*+}\pi^{-}$. In order to resolve the $D_1^0$ state and improve the
signal to background ratio,
the $D_2^{*0}$ state was suppressed by requiring $|\cos\alpha|>0.8$.
The $M(D^{*+}\pi^{-})-M(D^{*+})$ spectrum for all combinations passing the
above cuts is shown in Fig.\ \ref{figm1}(a). The prominent peak observed
is due to the $D_1^0$ state and the shoulder at higher mass is due to the
$D_2^{*0}$ state. This spectrum was fitted with a
fourth-order Chebychev polynomial for the background and two
Breit-Wigner resonance shapes convoluted
with Gaussian resolution functions for the signals. The $\sigma$'s of these
Gaussian functions were fixed to 4.0 MeV/c$^{2}$, as determined from Monte
Carlo studies. The mass and width of the higher mass convoluted
Breit-Wigner were
constrained to the measured values obtained above from the decay
$D_2^{*0} \rightarrow D^{+}\pi^{-}$, while the parameters of the other
convoluted Breit-Wigner were left free. For the suppressed $D_2^{*0}$
state, we obtained $48 \pm 30$ signal events. For the $D_1^{0}$ state,
the fit yields $286^{+51}_{-46}$ signal events,
$M(D_1^0)-M(D^{*+})= 411^{+1+2}_{-2-2}$ MeV/c$^{2}$, and
$\Gamma=20^{+6+3}_{-5-3}$ MeV/c$^{2}$.
The measured mass-difference corresponds
to a $D_1^0$ mass of $2421^{+1+2}_{-2-2}$ MeV/c$^{2}$. The second
error is systematic and was estimated by varying the cuts, the mass and
width of the fixed state, the background parameterization, and the spin of
the Breit-Wigner shapes used to describe the signals.
Our results for the mass and
width of this state, along with previous measurements, are listed in
Table \ref{tabm1}.

The spectrum of the mass-difference, $M(D^{*+}\pi^{-})-M(D^{*+})$,
with no cut on the helicity angle is shown in Fig.\ \ref{figm1}(b).
A fit to the mass-difference distribution with no helicity angle cut
yielded $M(D_1^0)-M(D^{*+})= 412 \pm 1$ MeV/c$^{2}$ and
$\Gamma = 24^{+5}_{-4}$ MeV/c$^{2}$ in excellent agreement with those obtained
above.

\section{Fragmentation}

The momentum spectra of $D_2^*(2460)^0 \rightarrow D^{+}\pi^{-}$ and
of $D_1(2420)^0 \rightarrow D^{*+}\pi^{-}$
were also obtained by fitting the
observed mass-difference distribution in five $x_{p}$ bins from 0.5 to 1.
Fitting the Peterson fragmentation function\cite{r13}:
\begin{equation}
   \frac{dN}{dx_{p}} \; \propto \; \frac{1}{x_{p}[1-\frac{1}{x_{p}}-
                        \frac{\epsilon_{p}}{1-x_{p}}]^{2}}
\end{equation}
to the acceptance-corrected momentum spectra,
shown in Fig.\ \ref{figxp}(a) and (b),
we find $\epsilon_{p}=0.034^{+0.018+0.005}_{-0.011-0.005}$ and
$0.015^{+0.004+0.001}_{-0.003-0.001}$, respectively,
for the $D_2^*(2460)^0$ and the $D_1(2420)^0$.
The systematic error was estimated by varying the number and size of the
$x_{p}$ bins used, and by varying the mass and the width of the $D_J^{0}$
states. Both spectra are quite hard compared with those observed for
continuum production of $D$ and
$D^*$ mesons at these center of mass energies \cite{r14,r15}.

\section{Cross Sections and Production Ratios}

As will be discussed below, measurements of the production rates for
the $D_J^{0}$ states
are, in general, sensitive to the $D_J^{0}$ alignment, which is uncertain.
When the detector acceptance is flat in $\cos\theta_{\pi}$,
the dependence on the alignment can be removed either by including events
from the full range of $\cos\theta_{\pi}$ or by selecting events with
$\cos\theta_{\pi} > 0$.
Monte Carlo studies showed that the efficiency for reconstructing
$D_2^*(2460)^0 \rightarrow D^{+}\pi^{-}$ does not vary over the range
$-1 \leq \cos\theta_{\pi} \leq +1$, while that of
$D_J^0 \rightarrow D^{*+}\pi^{-}$
does not vary significantly over the range $\cos\theta_{\pi}>0$.
We accordingly removed the $\cos\theta_{\pi}$ cut for
the decay of $D_2^*(2460)^0 \rightarrow D^{+}\pi^{-}$ and required
$\cos\theta_{\pi}>0$ for the decays of $D_J^0 \rightarrow D^{*+}\pi^{-}$.
The measured yields for the decay
modes $D_2^*(2460)^0 \rightarrow D^{+}\pi^{-}$,
$D_2^*(2460)^0 \rightarrow D^{*+}\pi^{-}$, and
$D_1(2420)^0 \rightarrow D^{*+}\pi^{-}$ were then determined to be
$513 \pm 81$, $164 \pm 41$, and $536 \pm 43$ signal events, respectively.
Using the fragmentation functions to extrapolate to zero momentum,
and after correcting for the efficiencies and the relevant $D^{0}$, $D^{+}$,
and $D^{*+}$ branching ratios,  we have extracted the production
cross sections times the branching ratios for $x_{p} \geq 0$:
\begin{equation}
\sigma(e^{+}e^{-} \rightarrow D_J^0X) \cdot B(D_J^0).
\end{equation}

Our measurements, as well as those of ARGUS\cite{r04,r05}, are summarized
in Table \ref{tabcs}.
In these calculations we have used
the new CLEO II\cite{r16} measurements of the $D^{*+}$ branching ratios.
In making comparisons with previous measurements, we scaled the old numbers
to compensate for the increase of the $B(D^{*+} \rightarrow D^{0}\pi^{+})$.
The systematic errors include the uncertainties in
the yields of the $D_J^0$ states, the uncertainties in the branching ratios,
and the uncertainty in the extrapolation to $x_{p}=0$.

We also extracted the production ratios:
\begin{equation}
N(D_J^0 \rightarrow D^{+}\pi^{-})_{x_{p}>0.6} \; / \; N(D^{+})_{x_{p}>0.6}
\end{equation}
and
\begin{equation}
N(D_J^0 \rightarrow D^{*+}\pi^{-})_{x_{p}>0.6} \; / \; N(D^{*+})_{x_{p}>0.6}.
\end{equation}
Our measurements, as well as those of ARGUS\cite{r04} and
CLEO 1.5\cite{r07}, are summarized in Table \ref{tabpr}.

We then determined the ratio of the branching fractions of the $D_2^*(2460)^0$
state:
\begin{equation}
\frac{B[D_2^*(2460)^0 \rightarrow D^+\pi^-]}
        {B[D_2^*(2460)^0 \rightarrow D^{*+}\pi^-]}
             = 2.2 \pm 0.7 \pm 0.6 .
\end{equation}
ARGUS\cite{r05} and CLEO 1.5\cite{r07} have measured this quantity to be
$3.7 \pm 1.4 \pm 1.9$ and $2.8 \pm 1.0$, respectively.
Our new result agrees well with the HQET predictions discussed previously.

\section{Helicity Angular Distributions}

The various spin-parity hypotheses and the corresponding helicity
angle distributions, for the general case of decays to
$D^*\pi$, are listed in Table \ref{tabsp}.

A study of the helicity angular distribution, $\cos\alpha$, lends
support to the identification of these states as members of
the $j=3/2$ doublet.
To study the helicity angle distributions of
$D_2^*(2460)^0 \rightarrow D^{*+}\pi^{-}$ and
$D_1(2420)^0 \rightarrow D^{*+}\pi^{-}$, we fitted the
$M(D^{*+}\pi^{-})-M(D^{*+})$ mass-difference spectra in five bins of
$\cos\alpha$ from $-1$ to $+1$. The masses and widths
of both states were fixed to our measured values given above.

In the decay of the $D_2^*(2460)^0$, the $D^{*+}$ and the
$\pi^{-}$ are emitted in a relative D-wave. This requires the $D^{*+}$
to have a helicity of $\pm 1$. Thus the form of the helicity angular
distribution for the $D_2^*(2460)^0$ state will be $\sin^2\alpha$, independent
of the initial alignment of this state.
Monte Carlo studies showed that our
efficiency does not vary significantly over the  helicity angle.
The number of $D_2^*(2460)^0$ events obtained in each $\cos\alpha$ bin
is shown in Fig.\ \ref{figh}(a).

The general form for the joint decay angular distribution for the
$D_1(2420)^0 \rightarrow D^{*+}\pi^{-}$ is given below, Eq. (8).
When the detector acceptance is flat in $\cos\theta_{\pi}$,
integration over $\cos\theta_{\pi} $ in either the range
$-1 \leq \cos\theta_{\pi} \leq +1$, or in the range $\cos\theta_{\pi} > 0$,
will remove dependence of the $D_1(2420)^0$ helicity angular distribution
on the alignment of the $D_1(2420)^0$ state.
Monte Carlo studies showed that, in the $\cos\theta_{\pi} > 0$ range, our
efficiency does not vary significantly over the plane of the helicity angle
and the decay angle. The only case where the efficiency is a little lower,
by $\approx 10\%$, is when  $\cos\alpha$ is backward and $\cos\theta_{\pi}$
is forward. In this case, the momentum of the slow $\pi^+$ in the decay
$D^{*+} \rightarrow D^{0}\pi^{+}$ has its minimum value.
We thus required $\cos\theta_{\pi} > 0$ and corrected for the relative
efficiency of the point at $\cos\alpha<-0.6$, and included the uncertainty in
the alignment of the $D_1(2420)^0$ state
in the systematic error of the efficiency correction.
This yielded the number of $D_1(2420)^0$ events in each $\cos\alpha$ bin
as shown in Fig.\ \ref{figh}(b).
In addition to the statistical errors, the error bars include systematic
errors due to the uncertainties in the
yields and due to the efficiency correction for the point at
$\cos\alpha<-0.6$.

We evaluated the $\chi^{2}$ per degree of freedom, $\chi^{2}/N_{dof}$,
and confidence level for various hypotheses for the shape of these
distributions.
The results are listed in Table \ref{tabcl}.
For the $D_2^*(2460)^0$ state the $\sin^{2}\alpha$
hypothesis is preferred, although the isotropic hypothesis is also
acceptable. For the $D_1(2420)^0$ state, the $\cos^2\alpha$
hypothesis is excluded for the first time, at more than 99\% CL.
This excludes an alternative interpretation of this state as a radial
excitation of the $D^0$ with $J^P = 0^-$. Because the $\sin^2\alpha$
hypothesis is also excluded, the $\cos\alpha$ distribution, alone, restricts
the $D_1(2420)^0$ state to the quantum numbers $J^P = 1^+, 2^-, 3^+ \ldots$
Since it is difficult to produce $L>1$ in the fragmentation process,
$J^P = 1^+$ is the preferred assignment. Because the $D_1(2420)^0$
width is relatively small, this state is identified as the $P_{1}^{(3/2)+}$
and the $D_2^*(2460)^0$ as its doublet partner, the $P_{2}^{(3/2)+}$.

To compare with previous results, we also fit the distribution for the
$D_2^*(2460)^0$ decay to the form, $A(1+B\cos^{2}\alpha)$, which gave the
result, $B=-0.74^{+0.49}_{-0.38}$ with $\chi^{2}/N_{dof}=0.6/3$ and a
CL $=90.6\%$. Fitting the distribution for the $D_1(2420)^0$ to the same
functional form gave the result, $B=2.74^{+1.40}_{-0.93}$ with
$\chi^{2}/N_{dof}=2.2/3$ and a CL $=53.2\%$.

Assuming that the $D_1(2420)^0$ does indeed have $J^P = 1^+$, both S and D
wave amplitudes are allowed in its decay to $D^{*+}\pi^-$. The general
form of the $D_1(2420)^0 \rightarrow D^{*+}\pi^{-}$ angular distribution is
then:
\begin{small}
\begin{eqnarray}
 \frac{d N}{d\cos\alpha \ d\cos\theta_\pi} & \propto &
  \frac{\sin^2\alpha}{8}
  \left[(1+\cos^2\theta_\pi)+\rho_{00}(1-3\cos^2\theta_\pi)\right]
  \left\{2\Gamma_S + \Gamma_D + 2\sqrt{2\Gamma_S\Gamma_D}\cos\varphi \right\}
  \nonumber \\
                                                   &    +    &
  \frac{\cos^2\alpha}{2}
  \left[(1-\cos^2\theta_\pi)-\rho_{00}(1-3\cos^2\theta_\pi)\right]
  \left\{\Gamma_S + 2\Gamma_D - 2\sqrt{2\Gamma_S\Gamma_D}\cos\varphi \right\},
\end{eqnarray}
\end{small}where $\alpha$ is the helicity angle, $\theta_\pi$ is the decay
angle, $\Gamma_S$ is the S-wave partial width, $\Gamma_D$ is the D-wave
partial width, $\varphi$ is the relative phase of the two amplitudes, and
$\rho_{00}$ is the fraction of $D_1(2420)^0$ with helicity 0 in the lab frame.
Integrating over $\cos\theta_\pi$ from --1 to +1, or from 0 to +1, gives:
\begin{small}
\begin{equation}
 \frac{1}{N} \frac{d N}{d \cos\alpha} =
 \frac{1}{2}
 \left\{\mbox{R} \ + \ \left(1-\mbox{R}\right)
 \left[ \frac{1+3\cos^{2}\alpha}{2} \right]
 \ + \ \sqrt{2\mbox{R}(1-\mbox{R})}
 \cos\varphi\left[ 1-3\cos^{2}\alpha \right] \right\},
\end{equation}
\end{small}where R=$\Gamma_S/(\Gamma_S + \Gamma_D)$.
Had we not released the $\cos\theta_\pi$ cut or applied a $\cos\theta_\pi>0$
cut, the helicity angle distribution would also have depended on the
alignment of the initial state.
Previous analyses\cite{r05,r07,r08} did not remove this cut.
Once R and $\cos\varphi$ are specified, the shape of the expected
$\cos\alpha$ distribution is fixed, and one can then determine the
$\chi^{2}$ that the distribution fits the data points
in Fig.\ \ref{figh}(b). The shaded region in Fig.\ \ref{fighp} shows the
region of the R-$\cos\varphi$ plane which is allowed at the 90\% CL.
The allowed regions
fall into two categories. First, if R is small then all values of
$\cos\varphi$ are allowed. Second, if $\cos\varphi$ is negative, then a
large S-wave partial width is allowed.  This can result from mixing between
the two $1^+$ states and it may explain the difference between the measured
value of the ratio $\Gamma(D_2^{*0})/\Gamma(D_1^0) \sim 1$, and the
HQET\cite{r02,r11} prediction $\sim 3$.

\section{Conclusions}

In conclusion, we have observed the two charmed mesons of masses
$2465 \pm 3 \pm 3$ and $2421^{+1+2}_{-2-2}$ MeV/c$^{2}$,  and of widths
$28^{+8+6}_{-7-6}$ and $20^{+6+3}_{-5-3}$ MeV/c$^{2}$, respectively.
The observed helicity angular
distributions are in good agreement with expectations for the $L=1$
$c\overline{u}$ $D_2^*(2460)^0$ and $D_1(2420)^0$ states.
We have made the first measurement of
{\small $\Gamma_S/(\Gamma_S + \Gamma_D)$} as a function of the
relative phase of S and D wave amplitudes in the
$D_1(2420)^0 \rightarrow D^{*+}\pi^{-}$
decay. In addition, the measured widths of both states are relatively narrow,
consistent with the predictions for D-wave decays of these states.
We have also determined the ratio of the branching
fractions of the two decay modes of $D_2^*(2460)^0$ into $ D^{+}\pi^{-}$ and
$D^{*+}\pi^{-}$, obtaining a value of $2.2 \pm 0.7 \pm 0.6$ which
is consistent with the expectation from HQET.
Taken together, these results constitute strong evidence
for identifying these two $D_J^0$ states as the $P_{2}^{(3/2)+}$ and
$P_{1}^{(3/2)+}$ ($j=3/2$) doublet predicted by HQET.

\acknowledgments

We gratefully acknowledge the effort of the CESR staff in providing us with
excellent luminosity and running conditions.
J.P.A. and P.S.D. thank
the PYI program of the NSF, I.P.J.S. thanks the YI program of the NSF,
G.E. thanks the Heisenberg Foundation,
K.K.G., I.P.J.S., and T.S. thank the
%
%
 TNRLC,
K.K.G., H.N.N., J.D.R., T.S.  and H.Y. thank the
OJI program of DOE
and P.R. thanks the A.P. Sloan Foundation for
support.
This work was supported by the National Science Foundation and the
U.S. Dept. of Energy.


\begin{table}
\caption{$D_2^*(2460)^0$ mass and width.}
\label{tabm2}
\begin{tabular}{ c c c }
Experiment & Mass (MeV/c$^{2}$) & Width (MeV/c$^{2}$)\\
\hline
CLEO II   & $2465 \pm 3 \pm 3$ & $28^{+8+6}_{-7-6}$\\
CLEO 1.5 & $2461 \pm 3 \pm 1$ & $20^{+9+9}_{-12-10}$\\
ARGUS & $2455 \pm 3 \pm 5$ & $15^{+13+5}_{-10-10}$\\
E691 & $2459 \pm 3 \pm 2$ & $20 \pm 10 \pm 5$\\
E687 & $2453 \pm 3 \pm 2$ & $25 \pm 10 \pm 5$\\
\end{tabular}
\end{table}

\bigskip

\bigskip

\begin{table}
\caption{$D_1(2420)^0$ mass and width.}
\label{tabm1}
\begin{tabular}{ c c c }
Experiment & Mass (MeV/c$^{2}$) & Width (MeV/c$^{2}$)\\
\hline
CLEO II & $2421^{+1+2}_{-2-2}$ & $20^{+6+3}_{-5-3}$\\
CLEO 1.5 & $2428 \pm 3 \pm 2$ & $23^{+8+10}_{-6-4}$\\
ARGUS & $2414 \pm 2 \pm 5$ & $13^{+6+10}_{-6-5}$\\
E691 & $2428 \pm 8 \pm 5$ & $58 \pm 14 \pm 10$\\
E687 & $2422 \pm 2 \pm 2$ & $15 \pm 8 \pm 4$\\
\end{tabular}
\end{table}

\bigskip

\bigskip

\begin{table}
\caption
{Summary of $D_J^0$ cross sections times branching ratios:
        $\sigma(e^{+}e^{-} \rightarrow D_J^0X) \cdot B(D_J^0)$.}
\label{tabcs}
\begin{tabular}{ c c c }
Decay Mode &  CLEO II (pb) & ARGUS (pb)\\
\hline
$D_2^*(2460)^0 \rightarrow D^{+}\pi^{-}$
       & $21.4 \pm 3.4 \pm 4.2$ & $68 \pm 21 \pm 28$\\
$D_2^*(2460)^0 \rightarrow D^{*+}\pi^{-}$
       & $9.5 \pm 2.4 \pm 1.8$ & $19 \pm 4 \pm 6$\\
$D_1(2420)^0 \rightarrow D^{*+}\pi^{-}$
       & $28.5 \pm 2.3 \pm 3.6$ & $32 \pm 4 \pm 9$\\
\end{tabular}
\end{table}

\bigskip

\bigskip

\begin{table}
\caption
{Summary of $D_J^0$ production ratios:
$N(D_J^0 \rightarrow D^{(*)+}\pi^{-})_{x_{p}>0.6} \;
       / \; N(D^{(*)+})_{x_{p}>0.6}$.}
\label{tabpr}
\begin{tabular}{ c c c c }
Decay Mode &  CLEO II (\%) & CLEO 1.5 (\%) & ARGUS (\%)\\
\hline
$D_2^*(2460)^0 \rightarrow D^{+}\pi^{-}$
       & $4.7 \pm 0.7 \pm 0.7$ & $10^{+2+2}_{-2-1}$ & $11^{+4+5}_{-4-5}$\\
$D_2^*(2460)^0 \rightarrow D^{*+}\pi^{-}$
       & $2.1 \pm 0.5 \pm 0.4$ & $3.6^{+1.0+0.4}_{-1.0-0.8}$ & \\
$D_1(2420)^0 \rightarrow D^{*+}\pi^{-}$
       & $6.8 \pm 0.6 \pm 0.9$ & $8.9^{+1.1+0.5}_{-1.1-0.5}$ & \\
\end{tabular}
\end{table}

\bigskip

\bigskip

\begin{table}
\caption
{List of spin-parity hypotheses and the corresponding helicity angle
distributions. $A_{\lambda 0}$ is the amplitude to produce $D^*$ with helicity
$\lambda$.}
\label{tabsp}
\begin{tabular}{ c c }
Spin-Parity Hypothesis &  Angular Distribution\\
\hline
$0^+$ & forbidden\\
$0^-$ & $\cos^{2}\alpha$\\
$1^-, 2^+, 3^-, \ldots$ & $\sin^{2}\alpha$\\
$1^+, 2^-, 3^+, \ldots$ & $|A_{10}|^2\sin^{2}\alpha+|A_{00}|^2\cos^{2}\alpha$\\
\end{tabular}
\end{table}

\bigskip

\bigskip

\begin{table}
\caption{ $\chi^{2}$ and CL for various angular distributions.}
\label{tabcl}
\begin{tabular}{ c c c c }
State & Angular Distribution & $\chi^{2}/N_{dof}$ & CL \\
\hline
$D_2^*(2460)^0$ & $\sin^{2}\alpha$ &1.2/4  & 87.8\% \\
                 & isotropic & 2.5/4 & 64.5\% \\
 & & & \\
$D_1(2420)^0$ & $1+3\cos^{2}\alpha$ & 2.3/4 & 68.1\% \\
                 & isotropic & 23.9/4 & $83.6\times10^{-6}$ \\
                 & $\cos^{2}\alpha$ & 28.2/4 & $11.4\times10^{-6}$ \\
                 & $\sin^{2}\alpha$ & 93.2/4 & $27.5\times10^{-20}$
\end{tabular}
\end{table}


\begin{figure}
\vspace{8.0 cm}
\includegraphics{plot_2m_n.ps}
\caption{The $M(D^{+}\pi^{-})-M(D^{+})$ mass-difference distribution.}
\label{figm2}
\end{figure}

\begin{figure}
\vspace{12.0 cm}
\includegraphics{plot_1m_n.ps}
\caption{The $M(D^{*+}\pi^{-})-M(D^{*+})$ mass-difference distribution for
         (a) $|\cos\alpha| \geq 0.8$ and (b) $-1 \leq \cos\alpha \leq +1$ .}
\label{figm1}
\end{figure}

\newpage
\makebox{}

\begin{figure}
\vspace{15.0 cm}
\includegraphics{plot_frag_n.ps}
\caption{The momentum spectra of (a) $D_2^*(2460)^0 \rightarrow D^{+}\pi^{-}$
         and (b) $D_1(2420)^0 \rightarrow D^{*+}\pi^{-}$.}
\label{figxp}
\end{figure}

\newpage
\makebox{}

\begin{figure}[htb]
\vspace{9.0 cm}
\includegraphics{plot_ha_n.ps}
\includegraphics{plot_hb_n_err.ps}
\caption{The normalized helicity angular distributions
     for (a) $D_2^*(2460)^0$ decay and (b) $D_1(2420)^0$ decay.}
\label{figh}
\end{figure}

\begin{figure}[htb]
\vspace{9.0 cm}
\includegraphics{plot_kut.ps}
\caption{Plot of R $= \Gamma_S / (\Gamma_S + \Gamma_D)$
versus cosine of the relative phase of S and D wave amplitudes in the
$D_1(2420)^0$ decay. The shaded area represents the 90\% confidence
level allowed region.}
\label{fighp}
\end{figure}

\end{document}